\documentclass{PoS}
\PoS{PoS(LAT2005)302}

\usepackage{epsfig,multicol}

\newcommand\fverb{\setbox\pippobox=\hbox\bgroup\verb}
\newcommand\fverbdo{\egroup\medskip\noindent%
			\fbox{\unhbox\pippobox}\ }
\newcommand\fverbit{\egroup\item[\fbox{\unhbox\pippobox}]}
\newcommand{\Slash}[1]{\ooalign{\hfil/\hfil\crcr$#1$}}
\newbox\pippobox
\def\PTP{Prog. Theor. Phys.(Kyoto)}

\def\NPB{{Nucl. Phys.} {\bf B}}

\def\PRL{Phys. Rev. Lett.}
\def\PRD{{Phys. Rev.} D}

\def\CPC{Comput. Phys. Commun.}
\def\dis{\displaystyle}

\title{Color Confinement in lattice Landau gauge with unquenched Wilson and KS fermions}

\author{\speaker{Hideo Nakajima} and Sadataka Furui$^\dagger$\\
	$^*$ Dept. of Infor. Sci., Utsunomiya Univ., Utsunomiya, 320-8585 Japan\\
	E-mail: \email{nakajima@is.utsunomiya-u.ac.jp}\\
	$^\dagger$ School of Sci.\& Engr., Teikyo Univ., Utsunomiya, 320-8551 Japan\\
	E-mail: \email{furui@umb.teikyo-u.ac.jp}}

\abstract{The Kugo-Ojima confinement criterion is verified in the unquenched Landau 
gauge QCD simulation. The valence quark propagator of the Kogut-Susskind 
fermion with use of the fermion action including the Naik term and the staple contribution is calculated on MILC Asqtad unquenched gauge configurations, 
and it shows infrared suppression of the quark propagator.
}

\FullConference{XXIIIrd International Symposium on Lattice Field Theory\\
25-30 July 2005\\
Trinity College, Dublin, Ireland}
\ShortTitle{Color Confinement in lattice Landau gauge}

\begin{document} 

%\maketitle  IS IGNORED %%%%%%%%%%%

\section{Introduction}
In the lattice Landau gauge QCD simulation, we adopt two types of the gauge field definitions\cite{NF99}:

$\log U$ type:
$U_{x,\mu}=e^{A_{x,\mu}},\ A_{x,\mu}^{\dag}=-A_{x,\mu},$

$U$ linear type:
$A_{x,\mu}=\displaystyle{1\over 2}(U_{x,\mu}-{U_{x,\mu}}^{\dag})|_{trlp.}$,

\noindent where $_{trlp.}$ implies traceless part.
 ( $\displaystyle A_\mu(x)=i\sum_a {A_\mu}^a(x)\frac{\Lambda^a}{\sqrt 2}$,\ \ 
${\rm tr} \Lambda^a \Lambda^b=\delta^{ab}$)

The corresponding optimizing functions are

$\log U$ type: $F_U(g)=||A^g||^2=\sum_{x,\mu}{\rm tr}
 \left({{A^g}_{x,\mu}}^{\dag}A^g_{x,\mu}\right)$,

$U$ linear type: $F_U(g)=\sum_{x,\mu}{\rm tr}\left (2- (\ U^g_{x,\mu}+{U^g_{x,\mu}}^{\dag})\right).$

 Under infinitesimal gauge transformation
$g^{-1}\delta g=\epsilon$, its variation reads for either defintion as 
\[
\Delta F_U(g)=-2\langle \partial A^g|\epsilon\rangle+
\langle \epsilon|-\partial { D(U^g)}|\epsilon\rangle+\cdots,
\]
where the covariant derivativative $D_{\mu}(U)$ for two options reads 
commonly as
\[
D_{\mu}(U_{x,\mu})\phi=S(U_{x,\mu})\partial_\mu \phi+[A_{x,\mu},\bar \phi]
\]
where 
$
\partial_\mu \phi=\phi(x+\mu)-\phi(x)$, and 
$\bar \phi=\dis{\phi(x+\mu)+\phi(x)\over 2}
$
 Stationality of the optimizing function implies Landau gauge, the local minimum implies Gribov Region\cite{Gr} and the global minimum implies Fundamental modular(FM) region\cite{Zw}.

We performed simulations on quenched configurations tabulated in Table \ref{quench}\cite{FN03,FN04}
and unquenched configurations of JLQCD\cite{jlqcd}, CP-PACS\cite{cppacs}, Columbia University\cite{cu} and MILC\cite{milc} tabulated in Table 2%\ref{unquench}
\cite{FN05}.
\begin{table}[hb]
\begin{center}
\begin{tabular}{c|c|c|c|c}
$\beta$ & $1/a$(GeV)  & $L$ & $aL(fm)$ & definition of A\\
\hline
6 &  1.97 & 16 & 1.60 & U-{linear}/ $\log U$\\
  &       & 24 & 2.40 & U-{linear}/ $\log U$\\
  &       & 32 & 3.20 & U-{linear}/ $\log U$\\
\hline
6.4 &  3.66 & 32 & 1.72 & U-{linear}/ $\log U$\\
  &       & 48 & 2.59 & U-{linear}/ $\log U$\\
  &       & 56 & 3.02 & U-{linear}/ $\log U$\\
\hline
6.45 &  3.87 & 56 & 2.86& $\log U$\\
\end{tabular}\label{quench}
\caption{Configurations used in the quenched QCD simulation}
\end{center}
\end{table}

\begin{center}
\begin{table}[htb]
\begin{tabular}{c|c|c|c|c|c|c|c|c}
   &$\beta$ &$K_{sea}$ & $am^{VWI}_{ud}/am^{VWI}_{s}$& $N_f$& $1/a$(GeV)&$L_s$ & $L_t$ &$a L_s$(fm)\\
\hline
JLQCD &5.2&  0.1340  & 0.134 & 2 & 2.221 & 20 & 48&1.78\\
      &5.2&  0.1355 & 0.093  & 2 & 2.221 & 20 & 48&1.78\\
\hline
CP-PACS & 2.1&  0.1357  & 0.087 & 2 & 1.834 & 24 & 48&2.58\\
        & 2.1 & 0.1382  & 0.020 & 2 & 1.834 & 24 & 48&2.58\\
\hline
CU &5.415&      & 0.025 & 2& 1.140 & 16 & 32&2.77\\
   &5.7&      & 0.010 & 2& 2.1 & 16 & 32 &1.50\\
\hline
MILC$_c$ &6.83($\beta_{imp}$)&      & 0.040/0.050& 2+1 & 1.64 & 20 &64&2.41\\
       &6.76($\beta_{imp}$)&      & 0.007/0.050& 2+1 & 1.64 & 20 &64&2.41\\
\hline
MILC$_f$ &7.11($\beta_{imp}$)&      & 0.0124/0.031& 2+1 & 2.19 &28 & 96&2.52\\
       &7.09($\beta_{imp}$)&      & 0.0062/0.031& 2+1 & 2.19 &28 & 96&2.52\\
\hline
\end{tabular}\label{unquench}
\caption{Configurations used in the unquenched QCD simulation}
\end{table}
\end{center}

\section{The Kugo-Ojima theory}
The Kugo-Ojima confinement criterion\cite{KO} is given by
the fact that the parameter $c$ defined as $u^{ab}(0)=-\delta^{ab}c$ in the
eq(\ref{kugo}) becomes 1
\begin{equation}\label{kugo}
(\delta_{\mu\nu}-{q_\mu q_\nu\over q^2})u^{ab}(q^2)={1\over V}
\sum_{x,y} e^{-ip(x-y)}\langle  {\rm tr}\left({\Lambda^a}^{\dag}
D_\mu \displaystyle{1\over -\partial D}[A_\nu,\Lambda^b] \right)_{xy}\rangle.\nonumber
\end{equation}

The Zwanziger horizon condition\cite{Zw} coincides
with Kugo-Ojima criterion provided the covariant derivative approaches
the naive continuum limit, i.e., $e/d=1$\cite{NFY01}.
We observe that in the quenched simulation $c$ saturated at about 0.8, while in the unquenched simulation it is consistent with 1.
\begin{center}
\begin{table}[htb]
\begin{tabular}{c|c|c|c|c|c|c}
&$K_{sea}$ or $\beta$ & $c_x$     & $c_t$    &$c$ &  $e/d$        &    $h$     \\
\hline
JLQCD&$K_{sea}=$0.1340 & 0.89(9)  & 0.72(4) &0.85(11) & 0.9296(2) & -0.08(11)  \\
&$K_{sea}=$0.1355 & 1.01(22) & 0.67(5) &0.92(24) & 0.9340(1) & -0.01(24)  \\
\hline
CP-PACS&$K_{sea}=$0.1357 & 0.86(6)  & 0.76(4) &0.84(7) & 0.9388(1) & -0.10(6)  \\
&$K_{sea}=$0.1382 & 0.89(9) & 0.72(4) &0.85(11) & 0.9409(1) & -0.05(9)  \\
\hline
CU&$\beta=$5.415 & 0.84(7)  & 0.74(4) &0.81(8) & 0.9242(3) & -0.11(8)  \\
&$\beta=$5.7 & 0.95(26)  & 0.58(6) &0.86(28) & 0.9414(2) & -0.08(28)  \\
\hline
MILC$_c$&$\beta=$6.76 & 1.04(11)  & 0.74(3) &0.97(16) & 0.9325(1) & 0.03(16)  \\
&$\beta=$6.83 & 0.99(14)  & 0.75(3) &0.93(16) & 0.9339(1) &  -0.00(16) \\
\hline
MILC$_f$&$\beta=$7.09 & 1.06(13)  & 0.76(3) &0.99(17) & 0.9409(1) & 0.04(17)  \\
  &$\beta=$7.11 &1.05(13)   &  0.76(3)& 0.98(17) & 0.9412(1) &  0.04(17) \\
\hline
\end{tabular}
\caption{The Kugo-Ojima parameter for the polarization along the spacial directions $c_x$ and that along the time direction $c_t$ and the average $c$, trace divided by the dimension $e/d$, horizon function deviation $h$ of the unquenched Wilson fermion(JLQCD, CP-PACS), and KS fermion (MILC$_c$,CU,MILC$_f$).  The  $\log U$ definition of the gauge field is adopted. }
\end{table}
\end{center}

\section{Quark propagator}
In the unquenched lattice simulation with the improved KS fermion action, the MILC collaboration has replaced the link variables by fattening\cite{OT} 
\[
U_\mu(x)\to c_1 U_\mu(x)+\sum_\nu w_3 S^{(3)}_{\mu\nu}(x)+\cdots
\]
where $S^{(3)}_{\mu\nu}$ is the staple contribution
\[
S^{(3)}_{\mu\nu}(x)=U_\nu(x)U_\mu(x+\hat \nu)U^\dagger_\nu(x+\hat\mu)
\]
and added the Naik term which is a product of three link variables along one
direction. 

The gauge configurations of the MILC collaboration are produced by incorporating, larger staple $S^{(5)}_{\mu\nu\rho}$ and $S^{(7)}_{\mu\nu\rho\sigma}$ and the tadpole improvement factor, and the action is called "Asqtad" 
action. 

In the present calculation of valence quark propagator, we do not include all improvements of the "Asqtad" action but incorporate $S^{(3)}$ staple term and the Naik term, i.e. the propagator we measure is the same as that of \cite{OT}, and we put for Dirac operator $\Slash{D}+m$ as
\begin{eqnarray}
\Slash{D}(U)_{x,y}&=&\dis{1\over 2}\sum_{\mu=-4}^4 \eta_\mu(x)sign(\mu)
[(c_1 U_\mu(x)+w_3\sum_{\nu\ne \mu}S_{\mu\nu}^{(3)}(x))\delta_{y,x+\hat\mu}\nonumber\\
&&+c_3U_\mu(x) U_\mu(x+\hat \mu)U_\mu(x+2\hat\mu)\delta_{y,x+3\hat\mu}]\nonumber
\end{eqnarray}
where $w_3=9/64$, $c_1=9/32$ and $c_3=-1/24$, and $\eta_\mu(x)$ is given as
%\eqnb
\[
\eta_\mu(x)=(-1)^{\zeta^{(\mu)}x},\ \ \ \ 
\zeta^{(\mu)}_\nu=\left\{ \begin{array}{ll} 1 & \nu<\mu\\
                                            0 & {\rm otherwise}
\end{array}\right.
\]

For KS fermion, one defines translation invariant states as
\[
\chi_{p,\alpha}(x)=\dis{1\over \sqrt{V}}e^{ikx},\ \ \ \ k_\mu=p_\mu+\pi
\alpha_\mu
\]
where $p_\mu=2\pi m_\mu/L_\mu$ $(m_\mu=0,...,(L_\mu/2)-1)$, $\alpha_\mu=0,1$.
Then the Dirac operator in the tree level, $\Slash{D}(I)+m$, has a proper form in the above basis as
\[
\langle \chi_{p',\beta}|\Slash{D}(I)+m|\chi_{p,\alpha}\rangle
=[i\sum_\mu(\bar \gamma_\mu)_{\alpha\beta}(\frac{9}{8}\sin p_\mu-\frac{1}{24}\sin 3p_\mu)+m\bar\delta_{\alpha\beta}]\delta_{p'p}
\]
where the Dirac gamma matrices of KS fermions appear as
\[
(\bar\gamma_\mu)_{\alpha\beta}=(-1)^{\alpha_\mu}\bar\delta_{\alpha+\zeta^{(\mu)},\beta}\ \ \ \ {\rm and}\ \ \bar\delta_{\alpha\beta}=\prod_\mu \delta_{\alpha_\mu \beta_\mu}.
\]

\section{Calculation of the propagator}
The quark propagator is calculated by statistical average over 
Landau-gauge-fixed samples as
\[
S_{\alpha\beta}(p)=\left\langle \langle \chi_{p,\alpha}| 
\dis{1\over \Slash{D}(U)+m }|\chi_{p,\beta}\rangle \right\rangle 
\]
The inversion, ${1\over \Slash{D}(U)+m }$, is performed via conjugate gradient method after preconditioning as follows.

 We define the operator ${\mathcal M}=(I+\frac{1}{m}\Slash {D})$ with use of even- odd- sites decomposition
\[
{\mathcal M}=\left(
\begin{array}{cc}
 I& \frac{1}{m}{\Slash {D}}_{oe}\\
   \frac{1}{m}{\Slash {D}}_{eo} & I
\end{array}\right)
=I-L-U
\]
where lower triangle and upper triangle matrices, $L$ and $U$, should be properly understood.

Using the Eisenstat trick\cite{FFGLLK}, we define
\begin{eqnarray}
\tilde {\mathcal M}&=&(I-L)^{-1}{\mathcal M}(I-U)^{-1}=(I+L)(I-U-L)(I+U)\nonumber\\
&=&I-LU=\left(\begin{array}{cc} I& 0\\
                          0 & I-(\frac{1}{m})^2{\Slash {D}}_{eo}{\Slash{D}}_{oe}\end{array}\right)\nonumber
\end{eqnarray}
where $L^2=U^2=0$ understood.

We note that $\tilde {\mathcal M}$ is hermitian and the conjugate gradient method and/or BiCGstab method are applicable for its inversion. Thus for solution 
of the equation $m{\mathcal M}\phi=\rho$, we obtain that
\[
\phi=(I+U){\tilde {\mathcal M}}^{-1}(I+L){1\over m}\rho
\]

 For the right hand side, $\rho$, of the equation ${\mathcal M}\phi = \rho$,
we put matrix site field $\rho_x = \chi_{p,\beta}(x)I_{3\times 3}$ where $I_{3\times 3}$ is the unit color source matrix. Then calculation with use of matrix site field, $\phi$, $\langle \chi_{p,\alpha}|\phi\rangle$
yields the sample contribution to $S_{\alpha\beta}(p)$ which is color $3\times 3$ 
matrix.

 Our error estimate of the inversion calculation is
\[
\frac{\| {\mathcal M}\phi-\rho \|}{\|\rho \|}<{\rm a\ few\ per\ cent\ at\ most}
\]
where the used norm is maximum norm in the space of site, color and flavor, 
and the accuracy gets $10^{-1}$ higher if $L^2$ norm is used.

With use of momentum definition 
\[
q_\mu={9\over 8}\sin(p_\mu)-{1\over 24}\sin(3p_\mu)=
\sin{p_\mu}\left(1+{1\over 6}\sin^2(p_\mu)\right),
\]
one obtains that
\[
S_{\alpha\beta}(q)= Z_2(q)\frac{-i(\bar\gamma q)_{\alpha\beta}+M(q)\bar\delta_{\alpha\beta}}{q^2+M(q)^2}
\]
Since ${\rm tr} \bar\gamma_\mu=0$, trace over color and flavor yields
\begin{equation}
{\rm tr} S(q)=16N_c \frac{Z_2(q)M(q)}{q^2+M(q)^2}\equiv 16N_c{\mathcal B}(q)\nonumber
\end{equation}
On the other hand 
\begin{equation}
{\rm tr} (i\bar\gamma q S(q))
=16N_c q^2 \frac{Z_2(q)}{q^2+M(q)^2}\equiv 16N_c q^2{\mathcal A}(q)\nonumber
\end{equation}

The dynamical mass function of the quark is 
$\displaystyle M(q)=\frac{{\mathcal B}(q)}{{\mathcal A}(q)}$,
and the quark wave function renormalization is
$\displaystyle Z_2(q)=\frac{{\mathcal A}(q)^2q^2+{\mathcal B}(q)^2}{{\mathcal A}(q)}$.

\section{Discussion and conclusion}

We found that the Kugo-Ojima parameter is consistent with 1 in the MILC configuration.

The quark field renormalization $Z_2$ is found to be infrared suppressed\cite{bhlpwz,lat05}. 
Its implication to the Kugo-Ojima confinement criterion is under investigation. We observed further, 
the renormalization of $Z_2$ and that of the running coupling are correlated\cite{lat05}. 
On the extraction of the continuum limit  of the running coupling from compact lattice 
simulations, there is a warning from the Dyson-Schwinger approach\cite{fga}.

\smallskip
\leftline{\bf Acknowledgement}
 We acknowledge discussion with Tony Williams, Christian Fischer and Patrick Bowman.
This work is supported by the KEK supercomputing project 05-128. 
H.N. is supported by the JSPS grant in aid of scientific research in priority area No.13135210.


\begin{thebibliography}{999}

\bibitem{KO} T. Kugo and I. Ojima, {\PTP} {Suppl.} {\bf 66}, 1 (1979).%{\it Local Covariant Operator Formalism of Non-Abelian Gauge Theories and Quark Confinement Problem}
\bibitem{Gr} V.N. Gribov, {\NPB} {\bf 139}{1}{(1978)}.%{\it Quantization of Non-Abelian Gauge Theories},

\bibitem{Zw} D. Zwanziger, {\NPB} {\bf 412}{(1994)} {657}.%{\it Fundamental modular region, Boltzmann factor and area law in lattice theory},

\bibitem{bhlpwz} P.O. Bowman et al., {\PRD}{\bf 71}{(2005)}{054507},{\tt hep-lat/0402032}.%{\it Unquenched quark propagator in Landau gauge},

\bibitem{OT} K. Orginos and D. Toussaint, {\NPB}(Proc. Suppl.){\bf 73}{(1999)}{909}.%{\it Tests of Improved Kogut-Susskind Fermion Actions},

\bibitem{FFGLLK} S. Fischer, A. Frommer, U. Gla\"asser, Th. Lippert, G. Litzenh\"ofer and K. Schilling,   {\CPC} {\bf 98}{(1996)}{20}.%{\it A parallel SSOR preconditioner for lattice QCD},

\bibitem{NF99} H. Nakajima and S. Furui,  {\NPB}(Proc. Suppl.){\bf 73}{(1999)}{865}. %{\it The new definition of lattice gauge fields and the Landau gauge},

\bibitem{NFY01} H. Nakajima, S. Furui and A.Yamaguchi,  {\NPB}(Proc. Suppl.){\bf 94}{(2001)}{558}.%{\it NUmerical studies of confinement in the lattice Landau gauge},

\bibitem{FN03} S. Furui and H. Nakajima,  {\PRD}{\bf 69},{074505}{(2004)}, {\tt hep-lat/0305010}.%{\it Infrared features of the Landau gauge QCD}

\bibitem{FN04} S. Furui and H. Nakajima,  {\PRD}{\bf 70},{094504}{(2004)}, {\tt hep-lat/0403021}.%{\it What the Gribov copy tells about confinement and the theory of dynamical chiral symmetry braeking}

\bibitem{FN05} S. Furui and H. Nakajima, {\it Infrared features of the Kogut-Susskind fermion and the Wilson fermion in Lattice Landau Gauge QCD}, {\tt hep-lat/0503029}.

\bibitem{lat05} S. Furui and H. Nakajima, {\it The running coupling in lattice Landau gauge with unquenched Wilson fermion and KS fermion}, this proceedings(PoS(LAT2005)291).%

\bibitem{jlqcd} S.Aoki et al., (JLQCD collaboration), {\PRD}{\bf 65},{094507}{(2002)}; ibid {\PRD}{\bf 68},{054502}{(2003)}.
\bibitem{cppacs} A. Ali Khan et al., (CP-PACS collaboration), {\PRD}{\bf 65},{054505}{(2002)}; S. Aoki et al.,(CP-PACS collaboration), {\PRD}{\bf 60},{114508}{(1999)}.
\bibitem{cu} F.R. Brown et al.,{\PRL}{\bf 67},{1062}{(1991)}.
\bibitem{milc} C.W. Bernard et al., {\PRD}{\bf 64},{054506}{(2001)}, {\tt hep-lat/0104002}
\bibitem{fga} C.S. Fischer, B.Gruter and R. Alkofer, {\it Solving coupled Dyson-Schwinger equations on a compact manifold}, {\tt hep-ph/0506053}.
\end{thebibliography}
\end{document}